\documentclass[12pt]{article}

\usepackage{amsmath}
\usepackage{amsfonts}
\usepackage{amscd}
\usepackage{amssymb}
\usepackage{amsthm}
\usepackage{bm}
\usepackage{nicefrac}
\usepackage{wrapfig}

\usepackage[usenames,dvipsnames]{xcolor}

\usepackage{lineno}
\usepackage{graphicx}
\usepackage[labelfont=bf]{caption}
\usepackage[format=hang]{subcaption}

\usepackage[algoruled]{algorithm2e}
\usepackage{listings}
\usepackage{fancyvrb}
\fvset{fontsize=\normalsize}

\lstdefinestyle{mystyle}{
    commentstyle=\color{OliveGreen},
    keywordstyle=\color{BurntOrange},
    numberstyle=\tiny\color{black!60},
    stringstyle=\color{darkblue},
    basicstyle=\ttfamily,
    breakatwhitespace=false,
    breaklines=true,
    captionpos=b,
    keepspaces=true,
    numbers=left,
    numbersep=5pt,
    showspaces=false,
    showstringspaces=false,
    showtabs=false,
    tabsize=2
}
\usepackage{enumitem}

\usepackage{hyperref}
\usepackage[all]{hypcap}
\hypersetup{colorlinks,linktoc=all}
\hypersetup{citecolor=MidnightBlue}
\hypersetup{urlcolor=MidnightBlue}
\hypersetup{linkcolor=black}

\usepackage[nameinlink, capitalize]{cleveref}
\creflabelformat{equation}{#1#2#3}
\crefname{equation}{Eq.}{Eqs.}
\Crefname{equation}{Eq.}{Eqs.}

\usepackage
[acronym,smallcaps,nowarn,section,nogroupskip,nonumberlist]{glossaries}
\glsdisablehyper{}
\setglossarystyle{long3col}
\makeglossaries

\usepackage{blindtext}
\usepackage{framed}

\usepackage{booktabs}
\usepackage{multirow}
\usepackage{longtable}
\usepackage{etoolbox,siunitx}
\robustify\bfseries
\usepackage{appendix}
\Crefname{appsec}{appendix}{appendices}
\AtBeginEnvironment{appendices}{\crefalias{section}{appendix}}

\newcommand{\g}{\,\vert\,}

\newcommand{\E}[1]{\mathbb{E}\left[#1\right]}
\newcommand{\EE}[2]{\mathbb{E}_{#1}\left[#2\right]}

\newtheorem{theorem}{Theorem}

\newtheorem{remark}{Remark}

\newtheorem*{example*}{Example}

\input{preamble/format}
\usepackage[%
minnames=1,maxnames=99,maxcitenames=2,	
style=alphabetic,
doi=false,	
url=false,	
natbib,	
backend=bibtex,	
sorting=nyt	
]{biblatex}%

\DeclareFieldFormat{sentencecase}{\MakeSentenceCase*{#1}}	

\renewbibmacro*{title}{%
  \ifthenelse{\iffieldundef{title}\AND\iffieldundef{subtitle}}	
    {}	
    {\ifthenelse{\ifentrytype{article}\OR\ifentrytype{inbook}%
      \OR\ifentrytype{incollection}\OR\ifentrytype{inproceedings}%
      \OR\ifentrytype{inreference}}	
      {\printtext[title]{%
        \printfield[sentencecase]{title}%
        \setunit{\subtitlepunct}%
        \printfield[sentencecase]{subtitle}}}%
      {\printtext[title]{%
        \printfield[titlecase]{title}%
        \setunit{\subtitlepunct}%
        \printfield[titlecase]{subtitle}}}%
     \newunit}%
  \printfield{titleaddon}}

\AtEveryBibitem{%
\ifentrytype{article}{	
    \clearfield{url}%
    \clearfield{urldate}%
    \clearfield{eprint}	
    \clearfield{eid}	
}{}	
\ifentrytype{book}{	
    \clearfield{url}%
    \clearfield{urldate}%
}{}	
\ifentrytype{collection}{	
    \clearfield{url}%
    \clearfield{urldate}%
}{}	
\ifentrytype{incollection}{	
    \clearfield{url}%
    \clearfield{urldate}%
}{}	
}	

\AtEveryBibitem{	
    \clearfield{pages}	
    \clearfield{review}%
    \clearfield{series}
    \clearfield{volume}	
    \clearfield{month}	
    \clearfield{eprint}	
    \clearfield{isbn}	
    \clearfield{issn}	
    \clearlist{location}	
    \clearfield{series}	
    \clearlist{publisher}	
    \clearname{editor}	
}{}	

\usepackage{tikz}
\usepackage{pgfplots, amsmath,amsthm, amssymb}
\usepackage{colortbl}
\usepackage{slashbox}
\pgfplotsset{compat=newest}
\usepgfplotslibrary{fillbetween}
\pgfplotsset{compat=newest}
\usetikzlibrary{patterns, arrows, arrows.meta,calc,decorations.pathreplacing,positioning}

\renewcommand{\arraystretch}{1.5}

\tikzset{brace/.style={decorate, decoration={brace}},
	brace mirrored/.style={decorate, decoration={brace,mirror}},
}

\newcounter{brace}
\usepackage{rotating}
\usetikzlibrary{patterns, arrows, arrows.meta,calc,positioning}
 \usepackage{colortbl}
 \usepackage{hhline}
\definecolor{azure(colorwheel)}{rgb}{0.0, 0.5, 1.0}
\definecolor{mygreen}{rgb}{0.502,1,0.6}

\tikzstyle{rot} = [rotate=-90, anchor= east]
\tikzstyle{st1} = [anchor=south , inner sep=1.2mm]
\usepackage{lipsum}
\usepackage{xr}

\usepackage[affil-it]{authblk}
\title{On the Assumptions of Synthetic Control Methods}
\author[1]{Claudia Shi}
\author[2]{Dhanya Sridhar}
\author[1]{Vishal Misra}
 \author[1]{David M. Blei}
\affil[1]{Columbia University}
 \affil[2]{University of Montreal}
 \date{}
 
\begin{document}
\maketitle

\begin{abstract}
Synthetic control (SC) methods have been widely applied to estimate the causal effect of large-scale interventions, e.g.\ the state-wide effect of a change in policy.
The idea of synthetic controls is to approximate one unit's counterfactual outcomes using a weighted combination of some other units' observed outcomes.
The motivating question of this paper is: how does the SC strategy lead to valid causal inferences?
We address this question by re-formulating the causal inference problem targeted by SC with a more fine-grained model, where we change the unit of the analysis from ``large units" (e.g.\ states) to ``small units" (e.g.\ individuals in states).
Under this re-formulation, we derive sufficient conditions for the non-parametric causal identification of the causal effect.
We highlight two implications of the reformulation: (1) it clarifies where ``linearity" comes from, and how it falls naturally out of the more fine-grained and flexible model, and (2) it suggests new ways of using available data with SC methods for valid causal inference, in particular, new ways of selecting observations from which to estimate the counterfactual.

\end{abstract}

\section{Introduction}
Since their introduction \citep{abadie2003economic,
  abadie2010synthetic}, synthetic control (SC) methods have become
commonplace for estimating causal effects from observational studies
with panel data.

Consider the following example. In 1988 California implemented a
large-scale tobacco control program, which increased the tobacco tax
by 25 cents. \citet{abadie2010synthetic} uses SC to study the effect
of this program on the average cigarette consumption in California.
The dataset contains the annual per-capita cigarette sales across a number of states, where none of the states other than California implemented a similar tobacco program. This dataset is illustrated in \cref{tb:panel}.
\begin{table}[ht!]
	\setlength{\tabcolsep}{10pt}
	\renewcommand{\arraystretch}{1.2}
	\centering
	\begin{tabular}{|l|ccc|c|cc|}
		\multicolumn{1}{l}{}          & \begin{sideways}$-$1970\end{sideways} & \begin{sideways}$-$1971\end{sideways} & \multicolumn{1}{c}{\begin{sideways}$-$1972\end{sideways}} & \multicolumn{1}{c}{$\qquad$} & \begin{sideways}$-$1988\end{sideways} & \multicolumn{1}{c}{\begin{sideways}$-$1989\end{sideways}}  \\ 
		\hhline{----~--|}
		California                            & 123                                     & 121                                     & 123.5                                                         &                              & 90.1                                    & {\cellcolor{azure(colorwheel)!60}}82.4                             
		  \\ 
		\hhline{====~==|}
		Alabama                            & 89.8                                     & 95.4                                     & 101.1                                                         &                              & 112.1                                    &105.6                                                          \\
		Arkansas                           & 100.3                                     & 104.1                                     & 103.9                                                         & $\cdots$                     & 121.5                                    & 118.3                                                         \\
		Virginia                            & 124.3                                     & 128.4                                     & 137                                                         &                              & 129.5                                     & 122.5                                                         \\
		Wisconsin                            & 106.4                                    & 105.4                                    & 108.8                                                         &                              & 102.6                                     & 100.3                                                          \\
		\multicolumn{1}{|c|}{$\vdots$} & $\vdots$                              & $\vdots$                              & $\vdots$                                                  &                              & $\vdots$                              & $\vdots$                                                   \\
		Wyoming                           & 132.2                                     & 131.7                                     & 140.0                                                       &                              & 114.3	                                     & 111.4                                                          \\
		\cline{1-4}\cline{6-7}
	\end{tabular}
\caption{
We observe the annual per-capita cigarette sales in packs of different states from 1970 to 1989. 
The blue shaded cell reports California's per-capita sales under a tobacco program.
The remaining cells report other states' per-capita sales without a tobacco program. 
}\label{tb:panel}
\end{table}

We observe that smoking in California decreased after the tobacco
program. However, we do not know whether the decrease is caused by the program or by other causes. Thus, to assess this difference, our goal is to estimate California's \emph{counterfactual outcome}. What would the 1989 smoking rate of California had been if its tobacco program had \emph{not} been implemented?

SC is a method to solve this problem. It uses the data from before
1989---when neither California nor the other states had implemented
the tobacco program---to learn a model of California's smoking rate as a weighted combination of the other states' smoking rates.
SC then uses the fitted weights to estimate California's counterfactual smoking rate in 1989.

In the general terminology of SC, California is the \textit{target},
the other states are the \textit{donors}, and the tobacco program is
the \textit{intervention}. A typical application of SC involves
aggregated time series data, such as in \cref{tb:panel}, with one target unit and a number of donor units. These units are often ``large'' units, such as states \citep{abadie2010synthetic,
  bohn2014did,
  cunningham2018decriminalizing}, counties \citep{abadie2003economic},
and districts \citep{bifulco2017using}. The causal question is one
about the target's counterfactual, after the intervention.

To justify SC estimators, existing works often make parametric assumptions about the true data generating process (DGP) of the potential outcomes of the aggregated data. A common assumption is that the potential outcomes under control (no tobacco program) are generated according to a linear factor model with additive noise
\citep{bai2009panel}. Follow-up works develop different estimators
based on this assumption \citep{xu2017generalized,
  imbens2021controlling}.

The purpose of this paper is to investigate the assumptions behind SC. When is it suitable to assume a linear factor model, and why can we write the target outcome as a weighted combination of the donors? 
We will show how to recover the SC methodology, but without making explicit parametric assumptions about the DGP of the potential outcomes.

The key idea behind this analysis is to construct a more fine-grained model of potential outcomes, one where we change the unit of analysis from a ``large unit" to a ``small unit." In the example, this change means that the potential outcomes are defined by individuals instead of states. We assume that different states index different distributions of individuals, and the state-level outcome of interest (average smoking rate) is an average of the individual outcomes in the state.

Under this fine-grained model, the causal effect targeted by classic SC is the sub-population average treatment effect (SATE), averaged over the individuals. We will derive sufficient conditions for the non-parametric causal identification of the SATE, and we will see that it uses the SC strategy. That is, we will derive sufficient conditions for the existence of a set of donors and weights such that their weighted combination can be used to approximate the effect.

This identification result further
suggests new settings in which to use the SC estimators, new ways to define the donors, and good heuristics for selecting auxiliary covariates. In more detail, there are several implications of this way of deriving the SC methodology.

First, the SC literature usually assumes that the control potential outcomes are generated according to a linear factor model. With the fine-grained model, we will see that the linear factor model form needs not be assumed a priori. 
Rather, the factor form is a natural consequence of invariance assumptions across groups (states) and time; and linearity is a consequence of the fact that expectation is a linear operator. A practical implication of this perspective is that SC can be used even if the individual-level DGP is non-linear.

Second, the SC literature does not generally offer guidance about the selection of the donors or how the choice of donors might affect the corresponding estimate. When reasoning from the fine-grained model, we will see how causal identifiability depends directly on properties of the target and selected donors; it may be possible to write California as a weighted combination of New York and Nevada, but not as a (non-zero) weighted combination of New York, Nevada, and Florida. Practically, the identification results suggest how to best choose donors for good SC estimation, and show that SC does not require that donors belong to the same type of group-level data. For example, we may correctly use data from Chicago (a city) to approximate the counterfactual of California (a state).

Finally, the analysis here provides a general heuristic of deciding which auxiliary covariates are suitable to include and which are not. We will see that alternative measurements of the target outcomes (e.g.\ per-capita tobacco spending measured in USD) are, in general, suitable auxiliary covariates,  whereas summaries of group-level characteristics (e.g.\  average age in states) may be unsuitable auxiliary covariates.

\paragraph{Organization. }
The paper proceeds as follows. In \cref{subsec:related}, we briefly review related work. 
In \cref{sec:data}, we formulate the tobacco tax example using both classical SC and individual-level potential outcomes and introduce the corresponding estimands.
In \cref{subsec:overview}, we introduce the SC estimators and review the common assumptions made in the SC literature.
In \cref{subsec:fine-grained}, we introduce the fine-grained model.
In \cref{subsec:id}, we establish sufficient conditions for causal identification of the estimand using the SC strategy.
In \cref{subsec:implication}, we discuss implications of the reformulation and the identification result, how it points to new settings in which to use SC estimators  and new ways to define donors.
In \cref{sec:aux}, we use the identification result to reason about what are suitable auxiliary covariates.
In \cref{sec:empirical}, we study these implications using simulation studies and the tobacco tax dataset.
Finally, in \cref{sec:discussion}, we discuss limitations and future work.

\subsection{Related Work.}\label{subsec:related}

This paper builds on the literature pioneered by \citet{abadie2003economic, abadie2010synthetic}.
A large part of the literature is about novel estimators \citep{abadie2011bias, abadie2015comparative, wong2015synthetic, doudchenko2016balancing, xu2017generalized, amjad2018robust, ben2018augmented, abadie2019penalized, amjad2019mrsc,arkhangelsky2019synthetic, li2020statistical,agarwal2020synthetic, athey2021matrix, imbens2021controlling} and inference methods \citep{abadie2010synthetic, doudchenko2016balancing, ferman2017placebo, shaikh2019randomization, chernozhukov2021exact}.
See \citet{abadie2019using} for an excellent review. 
This paper complements the existing work, as it interrogates the assumptions made by many of these estimators and inference methods.

This paper contributes to the growing effort of providing causal interpretations for synthetic controls 
\citet{o2016estimating}  formalize and synthesize the common assumptions in various methods for panel data inference.
\citet{bottmer2021design} study SC estimators properties under a randomized experiment setup.
\citet{shi2021proximal} develop identification and inference theory for the SC methods by drawing insights from the proximal causal inference literature \citep{miao2018identifying}.
However, all of this existing literature performs its analysis with group-level aggregates as units. In contrast, 
this paper takes advantage of the nature of the group-level data and shows how SC assumptions can arise from reasoning about individual-level potential outcomes.

Finally, this paper contributes to the growing research on invariance and causality \citep{scholkopf2012causal, bareinboim2014transportability, peters2016causal, buhlmann2018invariance,lei2020conformal, scholkopf2021toward}. In particular, a key assumption in this paper is the independent causal mechanism principle \citep{peters2017elements}.

\section{Data and Problem Formulation}\label{sec:data}
In this section we define the observed data, the group-level potential outcomes that underlie classical SC, and the individual-level potential outcomes that we consider in this paper. We define the causal estimand of interest in both settings.
For expository purposes we omit the auxiliary covariates for now. We introduce them in \cref{sec:aux}.

\subsection{The Observed Data} 
We have a dataset that contains the average smoking rate of $j= 1, ..., J$ states for $t= 1, ..., T$ time periods.
Let $\mu^{obs}_{jt}$ denote the average smoking rate of state $j$ and time $t$.
The \emph{target} state is $j=1$, i.e.\ California.
It is the only state that levied tobacco taxes. 
The remaining states $j \geq 1$ are potential \emph{donors}, which did not impose tobacco taxes.
The intervention (the tobacco taxes) happened at time $T_0$. To simplify notation, we assume one post-intervention time period, $T=T_0 + 1$, though the analysis easily generalizes to more post-intervention time periods.
The number of time periods $T$ and the potential donors $J$ are fixed.

\subsection{Classical SC Potential Outcomes} 
SC estimators are usually developed under the potential outcomes framework for causal inference \citep{neyman1923application, rubin1974estimating}.
Each state is considered a unit. The potential outcomes for each unit at each time period are $(\mu_{jt}(0), \mu_{jt}(1))$.
These variables are the average smoking rates of state $j$ at time $t$, one in the world where state $j$ increased tobacco taxes, and one in the world where state $j$ did not increase tobacco taxes.
We assume the treatment is well-defined and there is no interference between the states \citep{rubin1980randomization}.

The observed outcomes are: 

\begin{equation}
\mu^{obs}_{jt}=
\begin{cases}
  \mu_{jt}(1) & \text{if } j=1 \text{ and } t = T \\
  \mu_{jt}(0) & \text{otherwise. } 
\end{cases}
\end{equation}
In other words, we observe each states' smoking rate under no tobacco taxes \emph{except} for California at time $T$, where we observe its smoking rate with the tax.

\subsection{Fine-Grained Potential Outcomes}\label{subsec:individual-po}
This paper considers a more fine-grained model, where we treat individuals as units and states as distributions of individuals.
The pair $(Y_{ijt}(1), Y_{ijt}(0))$ denotes the potential outcomes of individual $i$ in group $j$ at time $t$,
e.g.\ how many packs of cigarettes person $i$ in state $j$ consumed at time $t$, under increased tobacco taxes or not.
Note we never make observations at an individual level, only in aggregate, but still we will reason about these variables.

We also consider the variable $X_{ijt} \in \mathbb{Z}^D$. It is a vector of causes that contribute to individual $i$'s outcome, such as age, education level, or income.
The relationship between the causes $X_{ijt}$ and potential outcomes $(Y_{ijt}(0), Y_{ijt}(1))$ can be linear or non-linear, and can also change across time periods.
We emphasize that we \emph{will not} observe these individual-level causes, but their existence will be crucial in our reasoning about the assumptions of SC.

We assume that interventions are made at a group-level and individuals in each group comply with their group-level intervention, i.e.\ individuals in California do not go to Nevada to purchase tobacco and vice versa.
The observed group-level averages approximate expected individual-level potential outcomes,
\begin{equation}
\mu^{obs}_{jt} \approx
\begin{cases}
  \E{Y_{jt}(1)} & \text{ if } j=1 \text{ and } t = T  \\
 \E{Y_{jt}(0)}& \text{otherwise.} 
\end{cases}\label{eq:consistency}
\end{equation}
The expectation is taken over the individuals $i$.

\subsection{Causal Estimand}\label{subsec:estimand}
We are interested in the causal effect of the tobacco taxes on the average smoking of individuals in California at time period $T$.
Using the classical SC notation, this estimand is formally defined as the unit-specific treatment effect,
\begin{align}
\tau_T = \mu_{1T}(1) - \mu_{1T}(0).\label{eq:ite}    
\end{align}
Under the individual-level notation, the estimand is the sub-population average treatment effect, averaged over individuals in the target distribution,\looseness=-1
\begin{align}
\tau_T = \E{Y_{1T}(1) - Y_{1T}(0)}.\label{eq:ate}    
\end{align}

\begin{remark}\label{rm:ite}
Being precise about the causal estimand is important, because different estimands require different identifying assumptions.
\cref{eq:ite} is the unit-specific treatment effect (UTE). \footnote{The more common name for \cref{eq:ite} is the individual treatment effect, where the units are individuals. To avoid potential confusion, we use UTE instead of ITE.} 
In general, UTE is incredibly difficult, if not impossible, to identify \citep{hernan2010causal}. 
It involves strong parametric assumptions on the data generating mechanism and the distribution of the noise variable.
In contrast, the causal estimand in \cref{eq:ate} is the an average causal effect.
Causal identification of the average causal effect is easier, and in general, does not require parametric assumptions \citep{rosenbaum1983central, imbens2004nonparametric}.\looseness=-1
\end{remark}
\section{A Fine-Grained Model for SC}\label{sec:new}
In this section, we first review the classical SC approach to causally identify and estimate the estimand.
We then develop a fine-grained model for SC, one that makes several assumptions that will eventually lead to the non-parametric identification of the causal estimand.
Finally, we discuss the practical implications of reasoning about the fine-grained model, and the corresponding identification result.

\subsection{Classical Synthetic Controls}\label{subsec:overview}
The idea behind SC is to use the observed outcomes of the donor states, which did not pass a tobacco tax, to help estimate the counterfactual California outcome. 
Specifically, SC posits that there exists a donor set $D$ in the potential donor pool $J$ and a weight set $\{\beta_j\}_{j\in D}$, such that,
\begin{align}
\mu_{1t}(0) = \sum_{j \in D} \beta_j \mu_{jt}(0) \quad \forall t \leq T.\label{eq:existence}
\end{align}

The validity of SC usually relies on parametric assumptions about the control potential outcomes $\mu_{jt}(0)$.
A common assumption is that the control potential outcomes are generated according to a linear factor model plus noise \citep{bai2009panel},
\begin{align}
\mu_{jt}(0) = \lambda_t^T\gamma_j + \epsilon_{jt}.\label{eq:factor_model}
\end{align}
Here $\lambda_t \in \mathbb{R}^R$ is a time-specific vector of factors shared across different units and $\gamma_j \in \mathbb{R}^R$ are unobserved unit-specific factor loadings.
The factor size $R$ is usually assumed to be significantly smaller than the number of potential donors $J$ and the total time periods $T$.
The noise variable $\epsilon_{jt}$ is zero centered.\footnote{The original SC \citep{abadie2010synthetic} for the tobacco tax example assume a variant of the factor model in \cref{eq:factor_model}, which we discuss in \cref{sec:aux}. }

Classical SC uses group-level observations to estimate the weights in \cref{eq:existence}. Specifically, it fits the regularized least squares,
\begin{align}
\hat{\beta} = \min_{\hat{\beta}_{j \in D}} \sum^{T_0}_{t=1} \big(\mu^{obs}_{1t} - \sum_{j \in D} \mu^{obs}_{jt} \cdot \hat{\beta}_j \big)^2 + \Upsilon( \hat{\beta}),\label{eq:estimator}
\end{align}
where $\Upsilon(\hat{\beta})$ is a prior or regularizer. SC uses the learned weights $\hat{\beta}$ to estimate the counterfactual California at time $T$.

To ensure a unique set of weights, existing works place restrictions on $\hat{\beta}$. 
For example, the original SC estimator \citep{abadie2010synthetic} restricts the weights to be positive and add up to one. 
\citet{doudchenko2016balancing} proposes an elastic-net regularizer.
\citet{robbins2017framework} suggests an entropy penalty.

\subsection{Invariance Assumptions for the Fine-Grained Model}\label{subsec:fine-grained}
We now consider the fine-grained model, where we reason about an individual $i$, and treat each state $j$ as a distribution of individuals.
We will show how to recover the SC strategy without making an explicit parametric assumption on the data generating process of the potential outcomes.

The target estimand is the sub-population average treatment effect in \cref{eq:ate}. 
Since the expected outcome under intervention $\E{Y_{1T}(1)}$ can be trivially identified, the goal is to causally identify the control expected outcome $\E{Y_{1T}(0)}$ from the data distribution.

In this subsection, we discuss the invariance assumptions that will lead to causal identification. In the next subsection, we will complete the derivation of how to identify the expected counterfactual.

The first assumption is that of an independent causal mechanism (ICM) \citep{scholkopf2012causal, scholkopf2021toward}.
ICM is a principle that the conditional distribution of each
variable given its causes (i.e.\ its ``mechanism") does not inform or influence the
other conditional distributions \citep{scholkopf2012causal}. 

In this context, the assumption means that the causal mechanism of an individual's tobacco consumption $Y_{ijt}(0)$ is independent of the distribution of their causes $X_{ijt}$.
If we know all the potential causes of an individual's tobacco consumption, then the distribution of the control potential outcome is independent of which distribution (i.e.\ state) the individual is from. 
\begin{description}[leftmargin=0cm]
\item[A1. (Independent Causal Mechanism)]
Conditional on the causes $X$, the potential outcome $Y(0)$ is independent of the population distribution $j$. 
For population distribution $j$ at time $t \leq T$, the joint distribution of $Y(0)$ and $X$ is
\begin{equation}\label{eq:invariant_mechanism} 
P_{jt}(X, Y(0)) = P_{jt}(X)P_t(Y(0)\g X).
\end{equation}
\end{description}
The assumption says that the distribution of individual causes $X$ can vary across states and time,
but the conditional outcome $Y(0) \g X$ only varies by time.
For each time point $t$, if we know all the potential causes of an individual's smoking behavior, which state they are from does not provide any additional information about the distribution of their control potential outcome.

Using \cref{eq:invariant_mechanism}, we rewrite the expected counterfactual as
\begin{align}
\E{Y_{jt}(0)} = \sum_x \underbrace{\EE{t}{Y(0) \g X=x}}_{\lambda_t}P_{jt}(X=x), \label{eq:iterated_expectation}
\end{align}
where $\mathbb{E}_t$ denotes the expectation with respect to $P_t(Y(0) | X=x)$, a distribution that is invariant across states.

\cref{eq:iterated_expectation} is similar to the factor model in \cref{eq:factor_model}, where the vector of factors $\lambda_t$ is the vector of conditional expected outcomes, but notice we did not make any assumptions about the relationship between the causes $X_{ijt}$ and potential outcomes $Y_{ijt}(0)$.
Rather, the linearity in \cref{eq:iterated_expectation} comes from the independent causal mechanism and iterated expectation --- the group-level outcomes are the averages of individual-level outcomes.

The second assumption is one of stable distributions. 
Given the target and the \emph{selected donors}, i.e.\ donors that will be used to construct the SC,
we can further decompose the causes $X$ into causes that differentiate the target and the selected donors, and causes that are invariant.

\begin{description}[leftmargin=0cm]
\item[A2. (Stable Distributions)]
Decompose the causes into two subsets $X=\{U, S\}$. Let $S$ denote the subset that differentiates the target from the selected donors, i.e., its distribution in the target group is different from its distribution in the selected donor groups. We assume that, for all groups, the distribution of $S$ does not change for all time periods $t \leq T$,
\begin{align}
 P_{jt}(X) = P_j(S)P_t(U \g S).\label{eq:factorization}
\end{align}
\end{description} 

In other words, The subset $S$ contains causes that vary across states, but are invariant across time. 
The conditional distribution of $U$ varies by time but is invariant across states.
We call $S$ the \emph{minimal invariant set}. 

\begin{remark}\label{rm:MIS}
The minimal invariant set $S$ is determined by the choice of the selected donors. 
For example, if we choose New Zealand as a donor to California, then the set $S$ may be the same as $X$, because these two groups are very different.
If the donor is a ``twin California," then the minimal invariant set $S$ is an empty set.
\end{remark}

With these assumptions in hand, we use \cref{eq:iterated_expectation} and \cref{eq:factorization} to rewrite the expected counterfactual,
\begin{align}
\E{Y_{jt}(0)} = \sum_s \underbrace{\EE{t}{Y(0) \g S=s}}_{\lambda_t}  \underbrace{P_j(S=s)}_{\gamma_j}.\label{eq:factor_expectation}
\end{align}
Comparing \cref{eq:factor_expectation} with the factor model in \cref{eq:factor_model}, we can see that the conditional expectations are analogous to the time varying factors $\lambda_t$. 
The probabilities are analogous to the state specific factor loadings $\gamma_j$.
What this equation shows is that with the two invariance assumptions, the expected outcome is naturally expressed as a factor model.
Using the fine-grained model, we discover that the ``linearity" in SC comes from aggregation, and the factor model arises from the invariance assumptions A1 \& A2. 

\subsection{Causal Identifiability}\label{subsec:id}
While \cref{eq:factor_expectation} is similar to the factor model in \cref{eq:factor_model}, it does not guarantee causal identifiability.
The reason is that the cardinality of the minimal invariant set $S$ can be very large.
In particular, if the cardinality of $S$ is larger than the number of donors then we cannot write the target as a weighted combination of the donors.

Here, we establish sufficient conditions for the causal identifiability of the causal estimand using the SC strategy.
Causal identifiability is about \emph{whether we can} express a causal estimand as a parameter of the observed distributions \citep{pearl2000causality}. In classical SC, causal identifiability is assumed in \cref{eq:existence}, that is, there exist a set of donors and weights, such that the target's counterfactual can be approximated as a weighted combination of the donors' observed outcomes.\footnote{Note, that causal identifiability is different from model identifiability, which is about whether we can recover a unique set of parameters from data.}

To establish causal identifiability, we need to make two more assumptions about the data distribution.
\begin{description}[leftmargin=0cm]
\item[A3. (Sufficiently Similar Donors)]
Let $D$ be the set of donors used to construct the synthetic control and
$S$ be the minimal invariant set for the target and the selected donors.
The donors are sufficiently similar if 
the cardinality of the donor set is greater than or equal to the cardinality of the minimal invariant set, 
\begin{align}
|D| \geq |S|.    
\end{align}
\end{description} 
As discussed in \cref{rm:MIS}, the cardinality of the minimal invariant set $S$ is determined by the selected donors. If the selected donors are very different from the target, the cardinality of $S$ is large.
If they are similar, the cardinality of $S$ is small.

\begin{description}[leftmargin=0cm]
\item[A4. (Target Donors Overlap)]
Let $ \{s_1, ..., s_R\}$ be the support of $S$.
There exists at least one donor distribution $j$ in the selected donor set $D$, where $ P_j(S=s) > 0 $.
\end{description} 
A4 implies that every type of individual living in California with characteristics $S=s$ might also live in one of the selected donor states. 

A3 and A4 are two assumptions on the distribution of the minimal invariant set $S$.
While we do not directly observe $S$ in practice, we can still use it to reason about the differences among the population distributions, and discuss its influence on causal identifiability.

Finally, we derive sufficient conditions for non-parametric identification of the causal estimand using the SC strategy.
\begin{theorem}(Causal Identifiability)
\label{thm:id}
Assume the causal mechanism is independent (A1), the target and the selected donors are stable during the periods of investigation (A2), the selected donors are sufficiently similar to each other (A3), and there is overlap between the target and the selected donors (A4). 
Then there exists a set of weights $\{\beta_{j}\}_{j \in D}$, such that across all time periods, the target's counterfactual can be written as a weighted combination of the donors' outcomes,
\begin{equation}
\E{Y_{1t}(0)} = \sum_{j \in D}  \beta_{j} \E{Y_{jt}(0)} \quad  \forall t \leq T.\label{eq:id}
\end{equation}
\end{theorem}
The proof is in \cref{app:proofs}.

The conditions in \cref{thm:id} lead to the assumption (\cref{eq:existence}) commonly made in the literature: the existence of synthetic controls.
Note that we have arrived at \cref{eq:existence} without making any parametric assumptions about the mechanism generating individual-level outcomes.

\subsection{Implications}\label{subsec:implication}
We have presented a set of causal assumptions that justify the SC methodology. What are the implications of these results?

\paragraph{Mixing types of donors.}
The fine-grained potential outcomes model in \cref{subsec:individual-po} provides insights about what can be treated as a donor.
Classical SC typically treats a state as a unit, and chooses donors as other states.
For example, \citet{abadie2010synthetic} excluded the District of Columbia as a possible 
donor.
The fine-grained model implies that each donor need only be a group of individuals, and does not need to be the same type of group as the target.
A donor's influence on the SC estimator has to do with how different it is to the target and other donors, i.e., the cardinality of the minimally invariant set.
For example, in \cref{sec:empirical}, we will see that we can use districts or regions as potential donors to a target state.

\paragraph{SC with non-linear DGPs.}
Previous work on SC begins with a linear factor model assumption, such as the one in \cref{eq:factor_model} \citep{xu2017generalized}.
However, it is unclear whether the role of the parametric model is for causal identification, for statistical necessity, or for notational convenience.
As discussed in \cref{rm:ite}, because the estimand is the unit-specific treatment effect, a natural interpretation of \cref{eq:factor_model} is that it is an assumption on the data generating mechanism.
Assuming the true mechanism is linear can be an unrealistic assumption.

In contrast, using the fine-grained model, we explained why, fundamentally, linearity can be a reasonable assumption in SC.
We cast the estimand as an average effect over sub-populations, and SC as a population re-weighting algorithm.
It make clear that the linear factor model in \cref{eq:factor_model} encodes invariance assumptions for causal identifiability, and 
that the linearity in \cref{eq:factor_model}
arises
from expectation being a linear operator.
A practical implication of this perspective is that
the causal identification result holds even if the fine-grained model involves a nonlinear mechanism.
Thus SC estimators are valid in settings where the individual-level DGPs are nonlinear.

\paragraph{The Role of $S$ and its relations to the donors.}
Classical SC assumes 
the latent factors $\lambda_t$ in \cref{eq:factor_model} are 
fixed and low rank \citep{abadie2019using}.
Consequentially, we may be tempted to use information from all the donors.
We show that different donors can lead to different latent factors.
Specifically, in \cref{eq:factor_expectation}, we draw the analogy between the size of the factors and the cardinality of the minimal invariant set $S$.
We show that the minimal invariant set $S$ is \emph{determined} by the donors used to construct the synthetic control.
Different donor sets can lead to different minimal invariant sets and naively including additional donors may lead to the non-existence of synthetic controls. 
Whether the assumption in \cref{eq:existence} holds is a property of the target and the selected donors.

\section{Auxiliary Covariates}\label{sec:aux}
So far, we have discussed how to analyze panel data as in \cref{tb:panel}.
In many practical settings, we may observe additional state-specific covariates.
For example, we may observe the percentage of teenagers in each state. 

Previous work incorporates the auxiliary covariates into the linear factor model \citep{abadie2015comparative}.
For example, \citet{abadie2010synthetic} posits the following model,
\begin{align}
\mu_{jt}(0) = \delta_t + \lambda_t^T\gamma_j +  \theta_t^TA_j + \epsilon_{jt},\label{eq:factor_aux}
\end{align}
where $\theta_t \in \mathbb{R}^R$ and $\delta_t \in \mathbb{R}^R$ are time-specific vectors of factors shared across different units and $A_j \in \mathbb{R}^k$ is a vector of $K$ observed state-specific covariates.

Specifically, \citet{abadie2010synthetic} assumes that is there exists a set of weights, such that, 
\begin{align}
A_{1k}  = \sum_{j \in D} \beta_{j} A_{jk} \quad \forall k \leq K  \quad
\text{and}\quad \mu_{1t}(0) = \sum_{j \in D} \beta_j \mu_{jt}(0) \quad \forall t \leq T.\label{eq:existence_aux}
\end{align}

Consequentially, one approach to using auxiliary covariates is to analyze them in parallel with the outcomes to solve for the SC weights \citep{abadie2010synthetic, abadie2015comparative, ben2018augmented, botosaru2019role}.

This use of auxiliary covariates requires us to assume \cref{eq:factor_aux}, that the underlying data generating process is linear.
A natural question is whether we can still use auxiliary covariates to construct the SC weights without assuming the underlying DGP is linear.
Here, we will use the fine-grained model to reason about  \emph{suitable auxiliary covariates}, those that inherently satisfy \cref{eq:existence_aux}.

Using \cref{eq:factor_expectation}, we can reason about several types of suitable auxiliary covariates. 
The first type are the state-specific probabilities of the variables in the minimal invariant set $S$: $P_j(S=s)$.
We observe that in \cref{eq:factor_expectation}, the relationships between the probabilities and the outcomes are linear.
Therefore, the probabilities have the same relationship with one another that the outcomes have with each other.
For example, suppose we know that the variable ``age'' differentiates the target and the donor distributions, i.e.\ ``age" is in $S$, then the ``percentage of young adults" can be a suitable covariate.

Note that some group-level summaries of individual-level characteristics may not be suitable auxiliary covariates.
For example, consider the average age (within a state).
Since we \emph{do not} assume a linear relationship between the individual-level characteristics and the individual-level outcomes,
we can not expect linear relationships between the group-level summaries and the group-level outcomes.

The second type of suitable auxiliary covariates are different measurements of the target outcomes. 
With a bit of algebra, we can see that the SC weights in \cref{eq:id} are combinations of the state-specific probabilities on the minimal invariant set: $P_j(S=s)$.
Recall that the minimal invariant set $S$ is solely determined by the target and the selected donor and that they are not influenced by the outcome measurements.
Therefore, the SC weights should be invariant to different measurements of the outcome variable $Y$.
For example, per-capita tobacco spending measured in USD would be a suitable auxiliary covariate to the target outcome: average cigarette consumption measured in packs.

Similarly, variables that share the same causes as the target outcomes and satisfy assumption A1 are suitable auxiliary covariates.
For example, if we believe that the same set of causes influence ``drinking''
and ``smoking'', and none of the states implemented any policy for alcohol consumption during the periods of consideration (A1 holds),  then ``annual average beer consumption" would be a suitable auxiliary covariate to average cigarette consumption.

\section{Empirical Studies }\label{sec:empirical}
We use simulations and tobacco-program data to study this perspective on SC and the implications of the theory around the fine-grained model.\footnote{code is available at \href{https://github.com/claudiashi57/fine-grained-SC}{github.com/claudiashi57/fine-grained-SC}}
We find the following.
\begin{enumerate}
    \item There is no reason the donors need to come from the same type of group as the target. We can still recover meaningful causal estimates when using donors from a different type.
    \item The SC estimators are valid even when the individual-level causal mechanisms are nonlinear.
    \item The cardinality of the minimal invariant set $S$ is dependent on the donor and target choices and is crucial to whether the weights can be generalized to the counterfactual.
    \item We may improve SC estimates when we include suitable auxiliary covariates in the analysis. However, unsuitable covariates will bias the estimates.
\end{enumerate}

\paragraph{Simulations.}
Following the fine-grained model in \cref{sec:new}, 
we first generate individual-level data, then construct group-level summaries.
The individual-level covariates $X$ take $K=12$ values.
Individual-level outcomes are derived from a set of non-linear and time-varying functions.
We create one target group and $5$ donor groups. Each group has a different composition of individuals. The compositions do not change over time.

We consider $T$ time periods.
For each group, at each time period, we sample $2000$ individuals according to its population composition.
The group-level summaries, $\mu^{obs}_{jt}$, are the average outcomes of individuals in group $j$ at time point $t$.
The randomness is on an individual level, instead of a group level.\footnote{More simulation details are in \cref{app:simulation}}

We create two knobs in the simulation, $S$ and $T$. 
The variable $T$ denotes the number of time periods, corresponding to the number of data points when fitting and evaluating the SC estimator.
The set $S$ is the minimal invariant set.
We use $|S|$ to denote the cardinality of $S$, and specifically how much the target and selected donor distributions differ from each other. 
$|S|$ ranges from $0$ to $K$. If the donors are identical to the target $|S|=0$. 
If the donors are very different in all aspects of the population composition $|S|=K$.
We do not observe the minimal invariant set $S$ and its cardinality.
The SC estimators do not use $S$ or $|S|$.

\begin{figure}
\centering
\includegraphics[width=.7\linewidth]{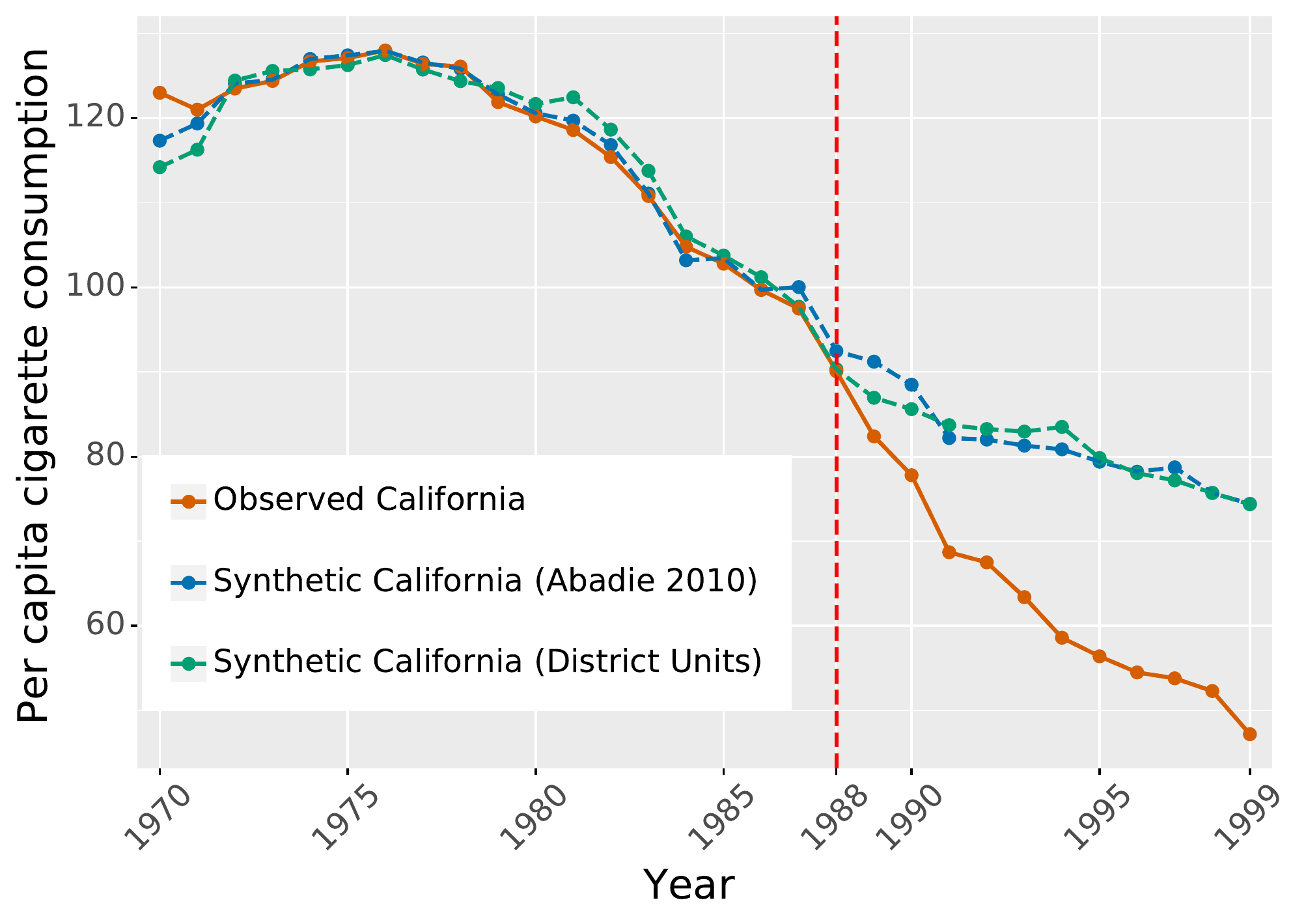}
\caption{Donors are not required to be of the same type: we use a linear combination of divisions to capture the outcome trends of California (a state) with high fidelity.}
\label{fig:prop99_regions}    
\end{figure}

\paragraph{Prop 99.}
Following \citet{abadie2010synthetic}, we analyze data about the Prop 99 tobacco program. We use the state-level data for the period 1970–2000.
We exclude states that also implemented large tobacco programs during the time frame and the states that raised tobacco tax by more than $50$ cents, resulting in $39$ potential donors. 
The outcome measurement is the per-capita cigarette sales in packs.
The main distinctions to \citet{abadie2010synthetic} are that (1) we include Washington DC in the donor pool, and (2) exclude the state-level covariates.

\paragraph{Methods and evaluation.}
For the Prop 99 example, we use the original SC estimator, with positive weights that sum up to one.
Since we cannot observe counterfactuals, we evaluate the estimation quality by plotting out the estimated counterfactuals.

For the simulation studies, we use ordinary least squares (OLS) as the SC estimator.
We use $75\%$ of the data to fit the estimator and $25\%$ of the data for out-of-sample evaluation.
The evaluation metric is the mean squared error averaged over data (time) points.
For expository purposes, when discussing the estimation quality, we use ``observed" and ``counterfactual" instead of ``in-sample" and ``out-of-sample".

\paragraph{(1) Mixing types of donors.}
As discussed in \cref{subsec:implication}, there is no reason to restrict donors to be of the same type as the target. 
To study this possibility empirically, we consider a type of donor that is different from states.
Since 1950, the United States Census Bureau has defined nine statistical divisions based on geographical location, e.g.\ New England, Mountain.
Each division contains several states.
We construct divisional-level donors using the United States census of 1990.
The average smoking rate of each division is a weighted combination of the smoking rate in its corresponding states, weighted by their population. 

We use the original SC estimator to construct a synthetic California using these divisional-level data.
We compare the counterfactual prediction of the divisional-level SC estimator with the original SC estimator.
As shown in \cref{fig:prop99_regions},  the synthetic California constructed by divisional-level data can capture the outcome trends of California with high-fidelity. 
We interpret the weights of the SC estimator in \cref{app:mixing}.

\begin{figure}
    \centering
    \includegraphics[width=.7\linewidth]{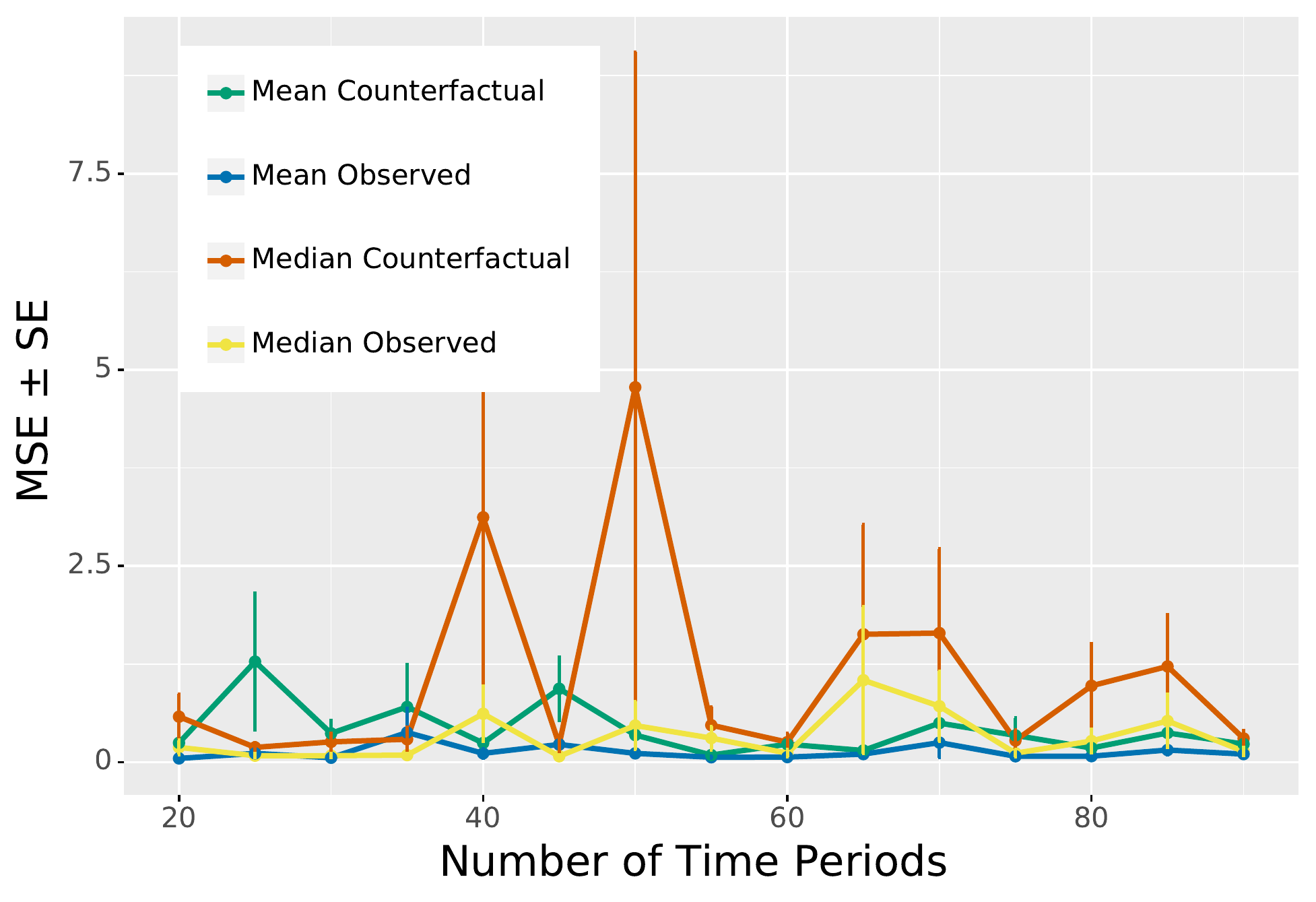}
    \caption{
    SC estimators are valid even when the
causal mechanisms generating individual-level outcomes are nonlinear.
 The figure reports the observed and counterfactual prediction loss, in settings where the group-level measurements are the ``mean'' or the ``median'' of the individuals in the groups. 
 }\label{fig:nonlinear}   
\end{figure}

\paragraph{(2) SC with no-linear DGPs.}
\cref{subsec:fine-grained} argues that the linearity in SC comes from aggregation, rather than a linear individual-level data generating process.
We study this claim empirically using the nonlinear data simulation described above.
We have data from six groups, one is the target and the other groups are the  donors.
We set the cardinality of $S$ to $5$ and consider a range of time periods for the data, from $T=20$ to $T=90$.

For a given number of time periods, we construct two panel datasets. 
The first dataset contains the \emph{average} outcomes of individuals in groups.
The second includes the \emph{median} outcomes of individuals in groups. 
Note that the ``mean'' is linear, where the ``median'' is nonlinear.

We apply the SC estimator to both panel datasets and evaluate the predictive performance on the observed and counterfactual data.
As shown in \cref{fig:nonlinear}, when the individual-level DGP is nonlinear, the SC estimator can still produce valid counterfactual estimates for the average outcomes.
Of course, linearity does not come for free. 
SC estimates are valid in this example because the ``mean" is a linear function, and the cardinality of the set $S$ is not greater than the number of available donors.
In contrast, when the measurement is the ``median", the SC estimator fails at predicting the counterfactual, because the median is a nonlinear function. Thus, we cannot expect a linear relationship between the median of the target group and the median of the donor groups.
\begin{figure}

\centering
\includegraphics[width=.7\linewidth]{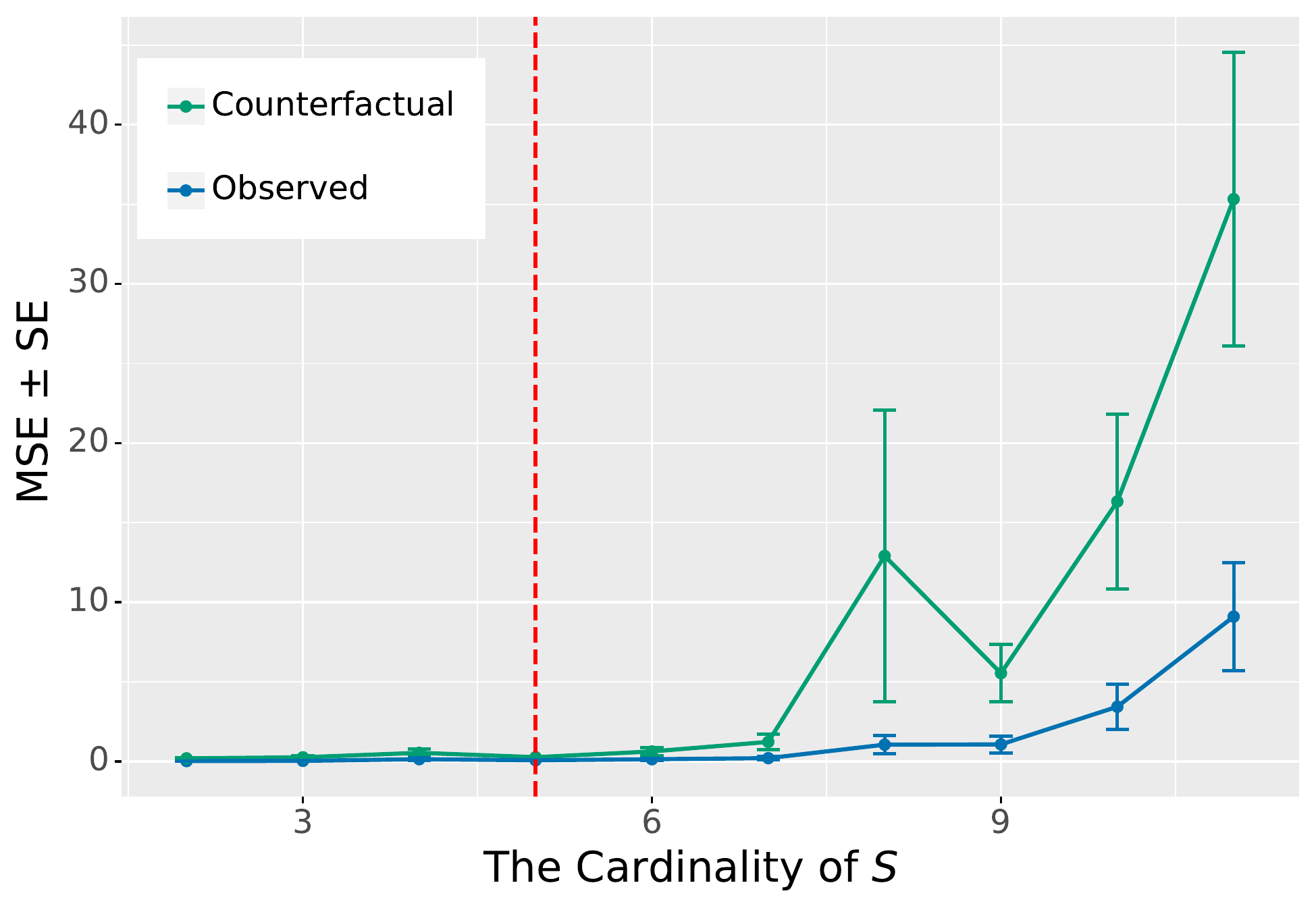}
\caption{The more different the donors are, the worse the SC counterfactual estimates become. The red dotted line denotes the number of donors used for constructing the synthetic controls.
}\label{fig:simulation_knobs}
\end{figure}

\paragraph{(3) The minimal invariant set $S$ is critical to causal identification.}
As discussed in \cref{subsec:id}, the minimal invariant set $S$ and whether \cref{eq:existence} holds are properties of the target and the donors.
If we choose donors that are drastically different from the target, the SC weights learned with the observed data may not generalize to the counterfactual data. We use nonlinear simulations to study the relationship between the donor choice and the quality of the counterfactual estimates. 
We fix the number of the donors to $5$, the number of time periods to $20$, and increase the cardinality of $S$ from $2$ to $11$.

As shown in \Cref{fig:simulation_knobs}, once the cardinality of $S$ surpasses the number of available donors, the counterfactual estimation error increases significantly.
Importantly, it is increasing at a significantly faster rate than the observed error.
We cannot determine whether the SC estimator can produce valid counterfactual estimates from the observed dataset alone.

\begin{table}[h]
\centering
\renewcommand{\tabcolsep}{3pt}
\captionsetup{width=\linewidth}

\caption{Using suitable covariates improves the estimation quality, whereas unsuitable covariates hurt the estimation quality. The table reports the mean squared error and standard error over 100 simulations.}\label{tb:estimation}
\begin{tabular}{l|cc}
MSE $\pm$ SE  & {\sc \textbf{Observed}} &  {\sc\textbf{Counterfactual} }\\
\hline
Outcome Only &$.07\pm .03$  & $.14\pm .05$  \\
Suitable Covariates & $.06 \pm .02$ & $.13\pm .05$ \\
Unsuitable Covariates & $.06\pm .02$ & $.24\pm .13$ \\
\hline
\end{tabular}
\end{table}

\paragraph{(4) Auxiliary Covariates.}
Finally, we study what happens to the counterfactual estimates when we include auxiliary covariates.
Using the simulations, we construct $10$ suitable auxiliary covariates and $10$ unsuitable auxiliary covariates.
The suitable covariates are the averages of the sine transformation of individual-level outcomes.
The unsuitable covariates are averages of the individual-level covariates.
We fix the number of the donors to $5$, the cardinality of the minimal invariant set to $5$, and the number of time periods to $15$.
We examine the observed and counterfactual estimation quality when including suitable covariates and including unsuitable covariates. 
As shown in \cref{tb:estimation}, incorporating suitable covariates may improve the counterfactual estimation, 
whereas including unsuitable covariates hurts the counterfactual estimation.
Notably, we may not infer whether an auxiliary covariate is suitable by looking only at the SC estimator's fit to the observed data.

\section{Discussion \& Future Work }\label{sec:discussion}
In this paper, we develop a fined-grained model for synthetic controls.
Using the tobacco example, we show that the ``linearity" in SC comes from aggregation: the group-level outcomes are averages of individual-level outcomes.
We further establish sufficient conditions for the non-parametric identification of the causal estimand and discuss several practical implications.

While this paper points to new ways of applying SC methods, the validity rests on the strong assumptions that an analyst must carefully consider.
For future work, we plan to establish uncertainty quantification methods that are compatible with the fine-grained model.
We also plan to develop sensitivity analyses that show how violations of the assumptions change the estimated effects.
\clearpage

\printbibliography
\clearpage

\makeatletter
\newtheorem{repeatthm@}{Theorem}
\newenvironment{repeatthm}[1]{%
    \def\therepeatthm@{\ref{#1}}
    \repeatthm@
}
{\endrepeatthm@}
\makeatother
\appendix

\begin{appendices}

\section{Proof For Thm 1}\label{app:proofs}
\begin{repeatthm}{thm:id}
Assume the target and the selected donors are stable during the periods of investigation (A2), the donors are sufficiently similar to each other (A3), and there is overlap between the target and the donors (A4). 
Then there exists a set of weights $\{\beta_{d}\}_{d \in D}$, such that across all time periods, the target's counterfactual can be written as a weighted combination of the donors' outcomes,
\begin{equation}
\E{Y_{1t}(0)} = \sum_{d \in D}  \beta_{d} \E{Y_{dt}(0)} \quad  \forall t \leq T.
\end{equation}
\end{repeatthm}
\begin{proof}
We first show that for a fixed time point $t$, there exist a set of weights such that the target can be written as a weighted combination of the donors.
We then show that there exists a set of weights that is invariant across time periods.

Fixing a time point $t$, we can write out \cref{eq:factor_expectation}, $\E{Y_{jt}(0)} = \sum_s \mathbb{E}_t{Y(0) \g S=s } P_j(S=s)$, for the target $j=1$ and the selected donors $D$.
Conceptually, we can think of the set expanded expectations as a system of linear equations, where the conditional expectations are the unknowns, and the probabilities are the scalars.
A4 says that the unknowns in the target equation are also in at least one of the donor equations.
A3 says that there are at least as many independent equations as the number of unknowns.
Combining A3 \& A4, we can ``solve" the unknowns.
Consequentially, we can write the target as a weighted combination of the donors, where the weights are functions of the probabilities.

A2 implies the probabilities are the same across time periods. Since the weights are functions of the probabilities, and the probabilities are invariant across time periods, the weights are also invariant across time periods. 
\end{proof}
\clearpage

\section{Simulation Details.}\label{app:simulation}
We discuss the details of the simulation studies.
Recall, the individual-level covariates take $K=12$ values.
There are $6$ groups. Each group has a different composition of individuals, denoted by $\pmb P_j$. 
The sparsity of the probabilities is determined by the cardinality of the minimal invariant set $S$.
\begin{align*}
\text{Group-level parameters}\\
\alpha_{sk} &\sim Bin(1,  1- \frac{|S|}{K})\\
\pmb \alpha_s &= (\alpha_{s1}, ..., \alpha_{sk})\\ 
\pmb P_j  & \sim Dir(K, \pmb \alpha_s)\\
\text{Individual-level data }\\
X_{ijt} &\sim Cat(K, \pmb P_j) \\
Y_{ijt} &  =f^t(X_{ijt}) + N(0,1)\\
\mu_{jt} & = \frac{1}{N_j} \sum Y_{ijt}
\end{align*}

Individual-level outcomes are derived from a set of non-linear and time-varying functions.
The exact functions are in the supplementary material.
The individual-level covariates take $K=12$ values.
At each point $t$ and for each group $j$, we sample $N_j=1000$ individuals.
The group-level summaries, $\mu^{obs}_{jt}$, are the average outcomes of individuals in group $j$ at time point $t$.

\clearpage
\section{Prop 99 Experiment Details.}\label{app:mixing}
In \cref{fig:prop99_regions}, we show that the synthetic California constructed by divisional level donors can capture the outcome trends of California with high fidelity. 
Here we describe how the donors are constructed.

Since 1950, the United States Census Bureau divided the United States into nine divisions: New England, Mid-Atlantic, East North Central, West North Central, South Atlantic, East South Central, West South Central, Mountain, and Pacific.
Each division consists of several states.
Following \citet{abadie2010synthetic}, we exclude states that also implemented large tobacco programs during the time frame and the states that raised tobacco tax by more than 50 cents: Massachusetts, Arizona, Oregon, Florida, Alaska, Hawaii, Maryland, Michigan, New Jersey, New York, and Washington.

We construct divisional-level donors according to the United States census of 1990. Each divisional donor is a weighted (by population) combination of its corresponding states that were not excluded for implementing similar policies.
We drop the Pacific division because all states in the division have been excluded.  
We use the original SC estimator to construct a synthetic California using these divisional-level data.

\begin{table}[ht]
  \begin{center}

    \begin{tabular}{l|r}
    \textbf{Division} & \textbf{Weight}\\
    \hline
     New England  & .05\\
     Mid-Atlantic & 0\\
     East North Central & 0 \\
     West North Central & 0\\
     South Atlantic & 0\\
     East South Central & 0\\
     West South Central & 0\\
     Mountain & .95\\
     Pacific & -\\
    \hline
    \end{tabular}
     \caption{Divisional-level weights used to construct the synthetic California. \label{tb:weights}}
  \end{center}
\end{table}

As shown in \cref{tb:weights},
the synthetic California uses only two donors: Mountain division and New England division. 
This is not surprising because California is geographically close to the Mountain division.
\Cref{tb:weights} is also consistent with the result in \citet{abadie2010synthetic}, where the chosen state-level donors come from either the Mountain division or the New England division.

Changing the unit of analysis does not necessarily harm interpretability. 
California is a state with a high population density.
According to the 1990 census, the population in California is around $30$ million.
The total population in the Mountain division used to construct the control is around $10$ million.
In contrast, the population in Montana, a donor used in \citet{abadie2010synthetic} with $0.2$ weights, is only around $0.8$ million.
Taking population level into consideration, it is as interpretable, if not more, to treat divisions as potential donors to California.

\end{appendices}

\end{document}
